\documentclass[preprint,prb]{revtex4}%
\usepackage{amssymb}
\usepackage{amsmath}
\usepackage{graphicx}%
\usepackage{amsfonts}
\providecommand{\U}[1]{\protect\rule{.1in}{.1in}}

\begin{document}
\title[ ]{Interaction of a Magnet and a Point Charge: Unrecognized Internal
Electromagnetic Momentum }
\author{Timothy H. Boyer}
\affiliation{Department of Physics, City College of the City University of New York, New
York, New York 10031}
\keywords{}
\pacs{}

\begin{abstract}
Whereas nonrelativistic mechanics always connects the total momentum of a
system to the motion of the center of mass, relativistic systems, such as
interacting electromagnetic charges, can have internal linear momentum in the
absence of motion of the center of energy of the system. \ This internal
linear momentum of the system is related to the controversial concept of
\textquotedblleft hidden momentum.\textquotedblright\ \ We suggest that the
term \textquotedblleft hidden momentum\textquotedblright\ be abandoned. \ Here
we use the relativistic conservation law for the center of energy to give an
unambiguous definition of the \textquotedblleft internal momentum of a
system,\textquotedblright\ and then we exhibit this internal momentum for the
system of a magnet (modeled as a circular ring of moving charges) and a
distant static point charge. \ The calculations provide clear illustrations of
this system for three cases: a) the moving charges of the magnet are assumed
to continue in their unperturbed motion, b) the moving charges of the magnet
are free to accelerate but have no mutual interactions, and c) the moving
charges of the magnet are free to accelerate and also interact with each
other. \ It is noted that when the current-carrying charges of the magnet are
allowed to interact, the magnet itself will contain internal
\textit{electromagnetic }linear momentum, something which has not been
presented clearly in the research and teaching literature.

\end{abstract}
\maketitle

\section{ Introduction}

\subsection{Problem of the Interaction of a Magnet and a Point Charge}

Although the idea of electromagnetic field momentum is more than a century
old, there is still confusion regarding this momentum. \ This confusion can be
found in connection with a magnet and a point charge. \ Now the interaction of
a magnet and point charge might seem to be so basic that it should be
well-understood in a junior-level course in electromagnetism. \ However, the
actual situation is quite the contrary. The momentum contained in stationary
distributions of electric charges and currents is still being debated in
teaching and research journals,\cite{resource} \ and the subject is tied up
with the controversy over what is termed \textquotedblleft hidden
momentum.\textquotedblright\cite{hid}\ \ In the present article, we will
discuss these questions yet again in the context of an unambiguous definition
of the \textquotedblleft internal momentum of a system\textquotedblright\ and
will then illustrate the concepts involved with a simple model for a magnet.

\subsection{Outline of the Presentation}

The first step in our presentation is a review of the fourth conservation law
involving either Galilean invariance or Lorentz invariance.\cite{B2005} \ (The
first three conservation laws involve energy, linear momentum, and angular
momentum.) \ We note that \textit{relativistic} theories allow the possibility
of internal momentum in a system which is unrelated to the motion of the
center of energy of the system. \ We use the fourth conservation law to give
an unambiguous definition of a system's internal momentum. \ We view the clear
idea of the \textquotedblleft internal momentum of a system\textquotedblright%
\ as a replacement for the ambiguous term \textquotedblleft hidden
momentum\textquotedblright\ which has provided the basis for an extended
controversy.\cite{hid}

The focus of our presentation involves internal momentum in the
electromagnetic system consisting of a magnet and a point charge. \ This is
the system where the concept of \textquotedblleft hidden
momentum\textquotedblright\ first arose. This simple system illustrates the
several basic variations of internal momentum. \ In addition, this system
reveals a form of internal electromagnetic linear momentum which does not seem
to be presented clearly in either the research or teaching literature.

We will use a point-charge model for the magnet. \ Thus our magnet can be
pictured as a set of point charges moving on a frictionless circular ring with
a point charge of opposite sign at the center so as to give neutrality to the
magnet. \ Now it appears that some physicists object to the use of a
point-charge model, and apparently insist that classical electromagnetism must
be discussed in terms of continuous charge and current distributions.
\ \ \ However, today, a century after Lorentz's classical electron theory, a
point-charge view seems natural, and such a view sometimes gives clarity to
difficult systems.

Next we outline the electromagnetic description provided by the Darwin
Lagrangian.\cite{DL} \ If we are interested in relativistic point charge
systems when radiation is not the subject of interest, then the natural
description of classical electrodynamics is that provided by the Darwin
Lagrangian. \ This badly-neglected approximation is frequently helpful; it has
been used in discussions of action and reaction between charge
particles,\cite{PA} Faraday induction,\cite{Faraday} mass-energy
equivalence,\cite{mass} and Lorentz transformations of relativistic energy and
momentum.\cite{Lorentz} \ Here it is enormously useful when discussing
internal momentum.

In the textbooks of electromagnetism, there are many examples and problems
where electric currents move in the presence of electrostatic
fields.\cite{G363},\cite{J286} \ A static point charge outside a solenoid or a
toroid is a typical such situation. However, usually there is no specification
as to the behavior of the charges which are carrying the currents. \ Are the
charges held to constant speed by external forces, or are the charges allowed
to accelerate but have no mutual interactions, or are the charges allowed to
accelerate and also have mutual interactions? \ Depending on the specific
assumption about the behavior of the current-carrying charges, there will be
different amounts of momentum in the system. \ In our examples in this
article, we describe these three different assumptions about the charged
particle behavior for the simplest situation which contains the basic physics,
the situation of a single current loop in the presence of a static point charge.

We start our illustrative calculations with the case in which the currents of
the magnet are held at their unperturbed value by nonelectromagnetic external
forces. \ This is the case which presumably is imagined in the literature,
though without any acknowledgement that external forces must be present to
hold the charges to their unperturbed motion. \ In this case, the familiar
electromagnetic momentum (due to the electric field of the external charge and
the magnetic field of the magnet) is an internal momentum for the system of
the magnet and the external point charge. \ We find that the external forces
acting on the moving charges of the magnet indeed provide the power flow
corresponding to the relativistic conservation law.

The next example removes the tangential nonelectromagnetic external forces
from the moving charges of the magnet but assumes that the moving magnet
charges do not interact with each other. \ In this case, since there are no
external forces\ providing local power to the magnet-point-charge system,
there can be no internal linear momentum for the system of the magnet and the
static charge. \ Since there is electromagnetic momentum present (associated
with the electric field of the external charge and the magnetic field of the
magnet), there must be an additional momentum in the system. \ Indeed, we
review the calculation for the relativistic \textit{mechanical} momentum in
the magnet. \ For the system of the magnet alone, this mechanical momentum is
an internal momentum; the electric field of the external charge provides the
external local power needed in the relativistic conservation law. \ This
internal mechanical momentum in the magnet is the familiar \textquotedblleft
hidden momentum\ in magnets\textquotedblright\ in the literature.\cite{SJ},
\cite{CV}

Finally, we consider the case where there are no nonelectromagnetic tangential
forces on the charges of the magnet so that these charges can accelerate, but
now the charges can also interact. \ For ease of calculation here, our example
involves only two interaction charges. \ Since we are averaging over the
initial orientation of the charges, this two-particle situation is equivalent
to many charges when each charge is allowed to interact only with one partner
charge on the opposite side of the circular current loop. \ (A more
sophisticated $N$-particle calculation, where there are mutual interactions
among all $N$ charges, is carried out in another manuscript.\cite{ABforces})
Once again, there is no internal momentum for the system of the magnet and the
point charge, but there must be internal momentum in the magnet. \ Here we
find that the internal momentum in the magnet is partially mechanical and
partially electromagnetic. \ In the multiparticle case when all the moving
charges interact with all the other charges, the internal momentum is
overwhelmingly electromagnetic, so that the internal \textit{mechanical}
momentum in the multiparticle magnet is negligible. \ \ A brief discussion
ends the article.

\section{The Relativistic Conservation Law and Internal Momentum in a System}

The notoriously difficult system of a magnet and a point charge is where
Shockley and James\cite{SJ} introduced the notion of \textquotedblleft hidden
momentum.\textquotedblright\ \ Here we wish to discuss this same system but to
use the idea of internal momentum.

Within nonrelativistic mechanics, all linear momentum is connected to particle
mass, $\mathbf{p=}m\mathbf{v.}$ \ The (time derivative of the) conservation
law associated with invariance under Galilean transformations is
\begin{equation}
\mathbf{P=}\left(
{\textstyle\sum\limits_{i}}
m_{i}\right)  \mathbf{V}_{c\text{of}m}\label{nonrel}%
\end{equation}
where $\mathbf{V}_{c\text{of}m}=d\mathbf{X}_{c\text{of}m}/dt$ and
$\mathbf{P}\,\ $is the total system momentum. \ Thus in nonrelativistic
mechanics, the total momentum of the system equals the total mass times the
velocity of the center of mass.

In contrast to nonrelativistic mechanics where all momentum is connected to
particle mass, electromagnetism contains linear momentum in the
electromagnetic fields, $\mathbf{P}_{em}=[1/(4\pi c)]\int d^{3}r\mathbf{E}%
\times\mathbf{B.}$ \ Since the magnetic field $\mathbf{B}$ is of order $1/c, $
the electromagnetic momentum $\mathbf{P}_{em}$ is of order $1/c^{2}$ and
corresponds to a relativistic effect.\cite{Gauss} \ The (time derivative of
the) conservation law associated with Lorentz transformations takes the
form\cite{B2005}%
\begin{equation}
\sum\limits_{i}\left(  \mathbf{F}_{\text{ext}i}\cdot\mathbf{v}_{i}\right)
\mathbf{r}_{i}=\frac{d}{dt}\left(  U\mathbf{X}_{c\text{of}E}\right)
-c^{2}\mathbf{P}\label{coE}%
\end{equation}
where $\mathbf{F}_{\text{ext}i}$ is the external force applied to the $i$th
particle of the system. \ This conservation law is sharply different from Eq.
(\ref{nonrel}) associated with Galilean invariance in nonrelativistic
mechanics. \ If we divide the relativistic equation (\ref{coE}) by $c^{2}$ and
take $c^{2}\rightarrow\infty,$ we indeed recover the nonrelativistic equation
(\ref{nonrel}). \ However, for a relativistic system, there is a new aspect;
the introduction of power $\mathbf{F}_{\text{ext}i}\cdot\mathbf{v}_{i}$ by an
external force $\mathbf{F}_{\text{ext}i}$ acting at position $\mathbf{r}%
\,_{i}$ leads to a changing position for the center of energy. \ This
situation has no counterpart in the nonrelativistic theory where the
introduction of power by an external force has no influence on the center of
mass apart from the motion of the masses.

For an isolated relativistic system where no external forces are present, the
system $U$ is constant and the relativistic conservation law (\ref{coE}) takes
the form%

\begin{equation}
\mathbf{P}=(U/c^{2})\mathbf{V}_{c\text{of}E}\label{PcoE}%
\end{equation}
so that the total momentum of the system is directly related to the velocity
of the center of energy. \ However, the relativistic conservation law
(\ref{coE}) now allows a new possibility. \ Now we can have system momentum
which is unrelated to the motion of the center of energy, but rather is
related to the local power introduced by external forces. \ In particular, one
can find situations where the total energy $U$\ and the center of energy
$\mathbf{X}_{c\text{of}E\text{ }}$ are not changing in time, $d(U\mathbf{X}%
_{c\text{of}E})/dt=0,$ but there is net linear momentum $\mathbf{P}_{int}$
given by%
\begin{equation}
\mathbf{P}_{int}=\frac{-1}{c^{2}}\sum\limits_{i}\left(  \mathbf{F}%
_{\text{ext}i}\cdot\mathbf{v}_{i}\right)  \mathbf{r}_{i}.\label{intm}%
\end{equation}
Here there is internal momentum in the system which is related to the local
delivery of power by the external forces. \ This last situation corresponds to
that where the presence of \textquotedblleft hidden momentum\textquotedblright%
\ has been claimed (by some authors) to be present in the system. \ We notice
from the factor of $1/c^{2}$ in Eq. (\ref{intm}) that we always expect the
internal momentum of a system to be of at least order $1/c^{2}.$ \ The term
\textquotedblleft hidden momentum,\textquotedblright\ was introduced by
Shockley and James\cite{SJ} in connection with the interaction of a magnet and
point charge. \ However, the term "hidden momentum" seems unfortunate, since
it suggests something mysterious. \ Furthermore, the use and misuse of the
designation has now made the term ambiguous.\cite{hid} Here we will not speak
of \textquotedblleft hidden momentum\textquotedblright\ but will refer to the
\textquotedblleft internal momentum in a system\textquotedblright%
\ $\mathbf{P}_{int}$ if there is linear momentum in the system which is
distinct from the motion of the system center of energy. \ With this
definition, a completely isolated system where no external forces are present
can contain no internal momentum. \ On the other hand, a subsystem of some
larger system can contain internal momentum in the subsystem due to the forces
arising from the other parts of the system; these forces from other parts of
the total system are external forces so far as the subsystem is concerned. \ \ 

\section{Basis for the Calculations}

\subsection{The Darwin Lagrangian}

We wish to examine the interaction of a magnet and a point charge through
relativistic order $v^{2}/c^{2}$ where electromagnetic field momentum appears.
\ Within classical electrodynamics, the electromagnetic interaction of point
charges $e_{i}$ at locations $\mathbf{r}_{i}$ moving with velocity
$\mathbf{v}_{i}$ can be described through order $v^{2}/c^{2}$ by the Darwin
Lagrangian\cite{DL}%

\begin{align}
\mathcal{L}  & \mathcal{=}%
{\textstyle\sum\limits_{i=1}^{i=N}}
m_{i}c^{2}\left(  -1+\frac{\mathbf{v}_{i}^{2}}{2c^{2}}+\frac{(\mathbf{v}%
_{i}^{2})^{2}}{8c^{4}}\right)  -\frac{1}{2}%
{\textstyle\sum\limits_{i=1}^{i=N}}
{\textstyle\sum\limits_{j\neq i}}
\frac{e_{i}e_{j}}{|\mathbf{r}_{i}-\mathbf{r}_{j}|}\nonumber\\
& +\frac{1}{2}%
{\textstyle\sum\limits_{i=1}^{i=N}}
{\textstyle\sum\limits_{j\neq i}}
\frac{e_{i}e_{j}}{2c^{2}}\left[  \frac{\mathbf{v}_{i}\cdot\mathbf{v}_{j}%
}{|\mathbf{r}_{i}-\mathbf{r}_{j}|}+\frac{\mathbf{v}_{i}\cdot(\mathbf{r}%
_{i}-\mathbf{r}_{j})\mathbf{v}_{j}\cdot(\mathbf{r}_{i}-\mathbf{r}_{j}%
)}{|\mathbf{r}_{i}-\mathbf{r}_{j}|^{3}}\right]  .\label{e1}%
\end{align}

\subsection{Electric and Magnetic Fields from the Darwin Lagrangian}

The Lagrangian equations of motion following from Eq. (\ref{e1}) can be
rewritten in the form of Newton's second law for the mechanical momentum
$\mathbf{p}^{mech}$ and force $\mathbf{F,}$ $d\mathbf{p}^{mech}/dt=d(m\gamma
\mathbf{v})/dt=\mathbf{F,}$ with $\gamma=(1-v^{2}/c^{2})^{-1/2}.$ \ In this
Newtonian form, we have%
\begin{align}
& \frac{d}{dt}\left[  m_{i}\gamma_{i}\mathbf{v}_{i}\right]  =\frac{d}%
{dt}\left[  \frac{m_{i}\mathbf{v}_{i}}{(1-\mathbf{v}_{i}^{2}/c^{2})^{1/2}%
}\right]  \approx\frac{d}{dt}\left[  m_{i}\left(  1+\frac{\mathbf{v}_{i}^{2}%
}{2c^{2}}\right)  \mathbf{v}_{i}\right] \nonumber\\
& =e_{i}%
{\textstyle\sum\limits_{j\neq i}}
\mathbf{E}_{j}(\mathbf{r}_{i},t)+e_{i}\frac{\mathbf{v}_{i}}{c}\times%
{\textstyle\sum\limits_{j\neq i}}
\mathbf{B}_{j}(\mathbf{r}_{i},t)\label{e5}%
\end{align}
with the Lorentz force on the $i$th particle arising from the electromagnetic
fields of the other particles. \ The electromagnetic fields due to the $j$th
particle are given through order $v^{2}/c^{2}$ by\cite{PA}%
\begin{align}
\mathbf{E}_{j}(\mathbf{r,t})  & =e_{j}\frac{(\mathbf{r}-\mathbf{r}_{j}%
)}{|\mathbf{r}-\mathbf{r}_{j}|^{3}}\left[  1+\frac{\mathbf{v}_{j}^{2}}{2c^{2}%
}-\frac{3}{2}\left(  \frac{\mathbf{v}_{j}\cdot(\mathbf{r}-\mathbf{r}_{j}%
)}{c|\mathbf{r}-\mathbf{r}_{j}|}\right)  ^{2}\right] \nonumber\\
& -\frac{e_{j}}{2c^{2}}\left(  \frac{\mathbf{a}_{j}}{|\mathbf{r}%
-\mathbf{r}_{j}|}+\frac{\mathbf{a}_{j}\cdot(\mathbf{r}-\mathbf{r}%
_{j})(\mathbf{r}-\mathbf{r}_{j})}{|\mathbf{r}-\mathbf{r}_{j}|^{3}}\right)
\label{e6}%
\end{align}
and
\begin{equation}
\mathbf{B}_{j}(\mathbf{r},t)=e_{j}\frac{\mathbf{v}_{j}}{c}\times
\frac{(\mathbf{r}-\mathbf{r}_{j})}{|\mathbf{r}-\mathbf{r}_{j}|^{3}}\label{e7}%
\end{equation}
where in Eq. (\ref{e6}) the quantity $\mathbf{a}_{j}$ refers to the
acceleration of the $j$th particle.

\subsection{Canonical Linear Momentum from the Darwin Lagrangian}

The canonical linear momentum $\mathbf{p}_{i}=\mathbf{p}_{i}^{mech}%
+\mathbf{p}_{i}^{field}$ associated with the $i$th charge is given by%
\begin{equation}
\frac{\partial\mathcal{L}}{\partial\mathbf{v}_{i}}=\mathbf{p}_{i}=m_{i}\left(
1+\frac{\mathbf{v}_{i}^{2}}{2c^{2}}\right)  \mathbf{v}_{i}+%
{\textstyle\sum\limits_{j\neq i}}
\frac{e_{i}e_{j}}{2c^{2}}\left(  \frac{\mathbf{v}_{j}}{|\mathbf{r}%
_{i}-\mathbf{r}_{j}|}+\frac{\mathbf{v}_{j}\cdot(\mathbf{r}_{i}-\mathbf{r}%
_{j})(\mathbf{r}_{i}-\mathbf{r}_{j})}{|\mathbf{r}_{i}-\mathbf{r}_{j}|^{3}%
}\right)  ,\label{e2}%
\end{equation}
corresponding to the \textit{mechanical} linear momentum $\mathbf{p}%
_{i}^{mech}$ of the $i$th particle%
\begin{equation}
\mathbf{p}_{i}^{mech}=m_{i}\gamma_{i}\mathbf{v}_{i}\approx m_{i}[1+v_{i}%
^{2}/(2c^{2})]\mathbf{v}_{i}=m_{i}\mathbf{v}_{i}+m_{i}v_{i}^{2}\mathbf{v}%
_{i}/(2c^{2}),\label{e2a}%
\end{equation}
and the \textit{electromagnetic} linear momenta $\mathbf{p}_{i}^{field}$
associated with the \textit{electric} field of the $i$th particle and
\textit{magnetic} fields of all the other $j$th particles\cite{J597}%
\begin{equation}
\mathbf{p}_{i}^{field}=%
{\displaystyle\sum\limits_{j\neq i}}
1/(4\pi c)%
{\textstyle\int}
d^{3}r\mathbf{E}_{i}\times\mathbf{B}_{j}=%
{\textstyle\sum\limits_{j\neq i}}
\frac{e_{i}e_{j}}{2c^{2}}\left(  \frac{\mathbf{v}_{j}}{|\mathbf{r}%
_{i}-\mathbf{r}_{j}|}+\frac{\mathbf{v}_{j}\cdot(\mathbf{r}_{i}-\mathbf{r}%
_{j})(\mathbf{r}_{i}-\mathbf{r}_{j})}{|\mathbf{r}_{i}-\mathbf{r}_{j}|^{3}%
}\right)  .\label{e2b}%
\end{equation}
\ Thus the total linear momentum $\mathbf{P}$ is given by
\begin{equation}
\mathbf{P}=%
{\displaystyle\sum\limits_{i}}
\mathbf{p}_{i}=%
{\textstyle\sum\limits_{i}}
m_{i}\left(  1+\frac{\mathbf{v}_{i}^{2}}{2c^{2}}\right)  \mathbf{v}_{i}%
+\frac{1}{2}%
{\textstyle\sum\limits_{i}}
{\textstyle\sum\limits_{j\neq i}}
\frac{e_{i}e_{j}}{2c^{2}}\left(  \frac{\mathbf{v}_{j}}{|\mathbf{r}%
_{i}-\mathbf{r}_{j}|}+\frac{\mathbf{v}_{j}\cdot(\mathbf{r}_{i}-\mathbf{r}%
_{j})(\mathbf{r}_{i}-\mathbf{r}_{j})}{|\mathbf{r}_{i}-\mathbf{r}_{j}|^{3}%
}\right)  .\label{e3}%
\end{equation}

The Darwin Lagrangian satisfies the familiar conservation laws involving
energy, linear momentum, and angular momentum. \ Through order $v^{2}/c^{2}$
the Darwin Lagrangian satisfies the fourth (and only specifically
relativistic) conservation law for relativistic systems involving the uniform
motion of the center of energy. \ Thus the Darwin Lagrangian allows us to
discuss systems which contain internal linear momentum in the sense of Eqs.
(\ref{coE}) and (\ref{intm}).

\section{Basic Model for the Magnet-Point-Charge Interaction}

\subsection{Magnet Modeled by Moving Point Charges}

The model for a magnet used in our calculations involves $N$ particles
$e_{i},$ each of charge $e$ and mass $m,$ which are held by external
centripetal forces of constraint in a circular orbit of radius $R$ centered on
the origin in the $xy$-plane while an opposite neutralizing charge $Q=$ $-Ne$
is located at the origin. \ The charges $e_{i}$ in the circular orbit are free
to move along the orbit and will accelerate due to any forces which are
tangential to the orbit. \ Thus in essence, the magnet is pictured as charged
beads $e$ sliding on a frictionless ring in the $xy$-plane with a balancing
opposite charge $Q$ located at the center of the ring. \ 

In the absence of any perturbing influence, the charges $e_{i}$ can be equally
spaced around the circular orbit with phases $\phi_{0i}=\omega_{0}t+\theta
_{i}$ where $\theta_{i}$ is an initial phase. \ The charges move with angular
velocity $\omega_{0}$, speed $v_{0}=\omega_{0}R,$ displacement%
\begin{equation}
\mathbf{r}_{0i}=R[\widehat{x}\cos(\omega_{0}t+\theta_{i})+\widehat{y}%
\sin(\omega_{0}t+\theta_{i})]=\widehat{r}_{0i}R,\label{n0}%
\end{equation}
and velocity%
\begin{equation}
\mathbf{v}_{0i}=\omega_{0}R[-\widehat{x}\sin(\omega_{0}t+\theta_{i}%
)+\widehat{y}\cos(\omega_{0}t+\theta_{i})]=\widehat{\phi}_{0i}\omega
_{0}R.\label{n00}%
\end{equation}
Here the radial and tangential unit vectors for the charge $e_{i}$ are
\begin{equation}
\widehat{r}_{0i}=\widehat{x}\cos(\omega_{0}t+\theta_{i})+\widehat{y}%
\sin(\omega_{0}t+\theta_{i})\label{n00a}%
\end{equation}
and%
\begin{equation}
\widehat{\phi}_{0i}=-\widehat{x}\sin(\omega_{0}t+\theta_{i})+\widehat{y}%
\cos(\omega_{0}t+\theta_{i}).\label{n00b}%
\end{equation}
The magnetic moment $\overrightarrow{\mu}$ of the $N$-moving-particle magnet
is given by the time-average of $e\mathbf{r\times v}/(2c)$ corresponding to
\begin{equation}
\overrightarrow{\mu}=\left\langle
{\textstyle\sum\limits_{i=1}^{N}}
e\mathbf{r}_{i}\mathbf{\times v}_{i}/(2c)\right\rangle =\widehat{z}%
NeR^{2}\omega_{0}/(2c).\label{e10}%
\end{equation}

We now introduce an external charge $q$ located on the $x$-axis at coordinate
$x_{q}$, $\mathbf{r}_{q}=\widehat{x}x_{q},$ which is held in place by
nonelectromagnetic external forces of constraint$.$ \ If the charge $q$ is far
away from the magnet, $R<<x_{q},$ then the electric field $\mathbf{E}%
_{q}(\mathbf{r})$ near the position of the magnet, $r\approx R<<x_{q},$ is
given by
\begin{equation}
\mathbf{E}_{q}(\mathbf{r})=\frac{q(\mathbf{r-r}_{q})}{\left\vert
\mathbf{r-r}_{q}\right\vert ^{3}}\approx\frac{q(\mathbf{r-}\widehat{x}x_{q}%
)}{x_{q}^{3}}\left(  1+\frac{3\widehat{x}\cdot\mathbf{r}}{x_{q}}\right)
=\frac{-\widehat{x}q}{x_{q}^{2}}+\frac{q[\mathbf{r-}3\widehat{x}%
(\widehat{x}\cdot\mathbf{r)]}}{x_{q}^{3}}\approx\frac{-\widehat{x}q}{x_{q}%
^{2}}\label{e11}%
\end{equation}
to order $R/x_{q}.$ \ In this article, we will need only the leading term in
the electric field, $\mathbf{E}_{q}(0)\approx-\widehat{x}q/x_{q}^{2}.$

We notice that for our model, the magnetic moment $\overrightarrow{\mu}$ in
Eq. (\ref{e10}) will remain unchanged provided that the average speed $\left(
%
{\textstyle\sum_{i}}
v_{i}\right)  /N$ of the the magnet charges remains unchanged. \ Since the
magnet charges move in a circular orbit in the $xy$-plane, $\mathbf{r}%
_{i}\mathbf{\times v}_{i}=\widehat{z}Rv_{i},$ and so the only possible change
in the magnetic moment is associated with a change in the average speed of the
magnet charges. \ In the calculations to follow, the average speed is
unchanged by the presence of the static charge $q$ and\ so the magnetic moment
is unchanged.

The central negative charge $Q$ will make no contribution to the momentum of
the magnet in our subsequent calculations since this charge is at rest at the
center of the circular ring. \ The canonical momentum $\mathbf{p}_{Q}$ of the
central charge $Q$ has no mechanical contribution because the charge is at
rest, and the electromagnetic field contribution vanishes because the moving
magnet charges have zero average velocity, which is unchanged by any
perturbation due to the external static charge $q$. \ Thus from Eq.
(\ref{e2}), we have%
\begin{equation}
\mathbf{p}_{Q}=M_{Q}0+%
{\textstyle\sum\limits_{i=1}^{N}}
\frac{Qe}{2c^{2}}\left(  \frac{\mathbf{v}_{i}}{R}+\frac{\mathbf{v}_{i}%
\cdot(\mathbf{r}_{i}-0)(\mathbf{r}_{i}-0)}{R^{3}}\right)  =0\label{s1}%
\end{equation}
since $\mathbf{v}_{i}\cdot\mathbf{r}_{i}=0$ and in all our calculations $%
{\textstyle\sum_{i}}
\mathbf{v}_{i}=0.$

\section{Magnet-Point-Charge Interaction for Unperturbed Magnet Currents}

\subsection{Familiar Electromagnetic Field Momentum for an Unperturbed Current
Loop in an External Electric Field}

In the approximation that the charges $e_{i}$ of the magnet are not perturbed
by the presence of the point charge $q$, we can obtain the linear momentum in
the electromagnetic field as the contribution of the electric field of the
point charge $q$ and the magnetic field arising from the moving magnet charges
$e_{i}.$ \ This corresponds to the canonical momentum of the stationary charge
$q$ (which is entirely electromagnetic field momentum),%
\begin{align}
\mathbf{p}_{q}  & =\mathbf{p}_{q}^{field}=%
{\textstyle\sum\limits_{i=1}^{N}}
\frac{1}{4\pi c}%
{\textstyle\int}
d^{3}r\mathbf{E}_{q}\times\mathbf{B}_{i}=%
{\textstyle\sum\limits_{j=1}^{N}}
\frac{qe}{2c^{2}}\left(  \frac{\mathbf{v}_{j}}{|\mathbf{r}_{q}-\mathbf{r}%
_{j}|}+\frac{\mathbf{v}_{j}\cdot(\mathbf{r}_{q}-\mathbf{r}_{j})(\mathbf{r}%
_{q}-\mathbf{r}_{j})}{|\mathbf{r}_{q}-\mathbf{r}_{j}|^{3}}\right) \nonumber\\
& \approx%
{\textstyle\sum\limits_{j=1}^{N}}
\frac{qe}{2c^{2}}\left[  \frac{\mathbf{v}_{j}}{x_{q}}\left(  1+\frac
{\widehat{x}\cdot\mathbf{r}_{j}}{x_{q}}\right)  +\frac{\mathbf{v}_{j}%
\cdot(\widehat{x}x_{q}-\mathbf{r}_{j})(\widehat{x}x_{q}-\mathbf{r}_{j})}%
{x_{q}^{3}}\left(  1+\frac{3\widehat{x}\cdot\mathbf{r}_{j}}{x_{q}}\right)
\right] \label{e12}%
\end{align}
\ We now insert the expressions (\ref{n0}) and (\ref{n00}) into Eq.
(\ref{e12}), average over time, and keep terms through first order
$1/x_{q}^{2}.$ \ We note that $\mathbf{v}_{j}\cdot\mathbf{r}_{j}=0,$
$\ \left\langle \mathbf{v}_{j}\right\rangle =0,$ $\ \left\langle
(\mathbf{v}_{j}\cdot\widehat{x})(\widehat{x}\cdot\mathbf{r}_{j})\right\rangle
=0,$ \ and \ $\left\langle \mathbf{v}_{j}(\widehat{x}\cdot\mathbf{r}%
_{j})\right\rangle =\widehat{y}\omega_{0}R^{2}/2=-\left\langle (\mathbf{v}%
_{j}\cdot\widehat{x})\mathbf{r}_{j}\right\rangle .$ \ The time-averaged
electromagnetic field momentum $\mathbf{p}_{q}$ receives equal contributions
from each charge, giving
\begin{equation}
\left\langle \mathbf{p}_{q}\right\rangle =\frac{\widehat{y}Nqe\omega_{0}R^{2}%
}{2c^{2}x_{q}^{2}}=\frac{1}{c}\mathbf{E}_{q}(0)\times\overrightarrow{\mu
}\label{e15}%
\end{equation}
This is the familiar expression for the electromagnetic field momentum for an
unperturbed localized steady current in an external electric field; it
corresponds to a problem in Jackson's text.\cite{J286} \ Even if the currents
of the magnet are perturbed by the electric field $\mathbf{E}_{q}(0) $, the
field momentum in Eq. (\ref{e15}) will remain unchanged to lowest order in
this perturbing field.

\subsection{Relativistic Conservation Law and Internal Momentum for
Unperturbed Currents in the Magnet}

In most treatments of the interaction a magnet and a point charge, it is not
pointed out that (in the presence of the external charge $q)$ the constant
currents of the magnet are made possible by nonelectromagnetic tangential
forces $\mathbf{F}_{\text{ext}i}$ on the charges $e_{i}$ carrying the magnet's
currents. \ These nonelectromagnetic external forces $\mathbf{F}_{\text{ext}%
i}$ on the charges $e_{i}$ balance the tangential force $e\mathbf{E}%
_{q}(\mathbf{r}_{i})$ of the electric field due to the external charge $q.$
\ Since these charges $e_{i}$ are moving in the electric field of the charge
$q$, the electric fields deliver local power $e\mathbf{E}_{q}(\mathbf{r}%
_{i})\cdot\mathbf{v}_{i}$ which must be absorbed by the nonelectromagnetic
forces $\mathbf{F}_{\text{ext}i}$ which keep the charges $e_{i}$ moving with
constant speed. \ Then the quantity on the left-hand side of the relativistic
conservation law in Eq. (\ref{coE}) and on the right-hand side of Eq.
(\ref{intm}) corresponds to
\begin{align}
\frac{1}{c^{2}}\sum\limits_{i}\left(  \mathbf{F}_{\text{ext}i}\cdot
\mathbf{v}_{i}\right)  \mathbf{r}_{i}  & =\frac{1}{c^{2}}\sum\limits_{i}%
\left(  -e\mathbf{E}_{q}(\mathbf{r}_{i})\cdot\mathbf{v}_{i}\right)
\mathbf{r}_{i}=\frac{1}{c^{2}}\sum\limits_{i}\left(  e\frac{\widehat{x}}%
{x_{q}^{2}}\cdot\mathbf{v}_{i}\right)  \mathbf{r}_{i}\nonumber\\
& =-\frac{\widehat{y}Nqe\omega_{0}R^{2}}{2c^{2}x_{q}^{2}}=-\left\langle
\mathbf{p}_{q}\right\rangle \label{ee16}%
\end{align}
where we have again used $\widehat{y}\omega_{0}R^{2}/2=-\left\langle
(\mathbf{v}_{j}\cdot\widehat{x})\mathbf{r}_{j}\right\rangle .$ \ Thus exactly
as expected, the internal momentum of the system of the point charge and the
magnet with unperturbed currents is non-zero, and this internal momentum
corresponds exactly to the electromagnetic field momentum involving the
electric field of the charge $q$ and the magnetic field of the magnet given in
Eq. (\ref{e15}). \ This familiar electromagnetic momentum is not usually
classified as "hidden momentum" in the muddled literature of "hidden
momentum." \ However, this momentum is exactly linear momentum which does not
involve motion of the center of energy of the system and is associated with
the introduction of local power as in the form (\ref{intm}) of the
relativistic conservation law. \ 

In this case, there is no internal momentum in the subsystem of the magnet
alone; the equally-spaced magnet particles move with constant speed in a
circle so $%
{\textstyle\sum_{i}}
\mathbf{p}_{i}=0$. \ This fits with the relativistic Eq. (\ref{intm}) since
there is no net local delivery of power by external forces on the magnet; the
electric forces $e\mathbf{E}_{q}(\mathbf{r}_{i})$ (which are now external
forces) on the magnetic charges $e_{i}$ due to the charge $q$ are balanced by
the nonelectromagnetic tangential forces $\mathbf{F}_{\text{ext}i}$, so that
the power delivered by the electric field of the charge $q$ is balanced by the
power absorbed by the nonelectromagnetic tangential forces.

\section{Magnet-Point-Charge Interaction for Freely-Moving Non-interacting
Charges in the Magnet}

\subsection{Perturbation of the One-Moving-Particle Magnet}

Our next model for the magnet currents involves $N$ \textit{non-interacting}
charges which are free to move in the direction tangential to the circular
orbit. \ Thus we may picture the charges as sliding freely on the frictionless
ring in response to the electric field of the external charge. \ Since the
charges have no mutual interactions, we may treat each charge separately as
though the magnet involved only one moving charge of mass $m$ and charge $e$.
\ For this one-moving-particle magnet, the index \textquotedblleft%
$i$\textquotedblright\ corresponds only to $i=1$. \ The neutralizing charge
$Q=-e$ is located at the origin. \ 

The position of the charged particle $e$ in the magnet is perturbed by the
presence of the external electric field due to the charge $q.$ The phase angle
is no longer the $\phi_{0i}=\omega_{0}t+\theta_{i}$ appearing in Eq.
(\ref{n0}), but rather becomes \ $\phi_{i}=\omega_{0}t+\theta_{i}+\eta
_{i}(t),$ so that the displacement of the particle through first order in the
perturbation is now
\begin{equation}
\mathbf{r}_{i}=R\{\widehat{x}\cos[\omega_{0}t+\theta_{i}+\eta_{i}%
(t)]+\widehat{y}\sin[\omega_{0}t+\theta_{i}+\eta_{i}(t)]\}=R\widehat{r}%
_{i}=\mathbf{r}_{0i}+\delta\mathbf{r}_{i}\label{n2}%
\end{equation}
where%
\begin{equation}
\widehat{r}_{i}=\widehat{x}\cos[\omega_{0}t+\theta_{i}+\eta_{i}%
(t)]+\widehat{y}\sin[\omega_{0}t+\theta_{i}+\eta_{i}(t)],\label{n2a}%
\end{equation}
the unperturbed displacement $\mathbf{r}_{0i}$ is given in Eq. (\ref{n0}),
and
\begin{equation}
\delta\mathbf{r}_{i}=\eta_{i}R\{-\widehat{x}\sin[\omega_{0}t+\theta
_{i}]+\widehat{y}\cos[\omega_{0}t+\theta_{i}]\}=\widehat{\phi}_{0i}R\eta
_{i}.\label{nn2}%
\end{equation}
Due to the perturbation, the velocity $\mathbf{v}_{i}=d\mathbf{r}_{i}/dt$ is
now%
\begin{align}
\mathbf{v}_{i}  & =(\omega_{0}+d\eta_{i}/dt)R\{-\widehat{x}\sin[\omega
_{0}t+\theta_{i}+\eta_{i}(t)]+\widehat{y}\cos[\omega_{0}t+\theta_{i}+\eta
_{i}(t)]\}\nonumber\\
& =\widehat{\phi}_{i}(\omega_{0}+d\eta_{i}/dt)R=\mathbf{v}_{0i}+\delta
\mathbf{v}_{i}\label{n3}%
\end{align}
where%
\begin{equation}
\widehat{\phi}_{i}=-\widehat{x}\sin[\omega_{0}t+\theta_{i}+\eta_{i}%
(t)]+\widehat{y}\cos[\omega_{0}t+\theta_{i}+\eta_{i}(t)],\label{n3a}%
\end{equation}
the unperturbed velocity $\mathbf{v}_{0i}$ is given in Eq. (\ref{n00}), and
\begin{equation}
\delta\mathbf{v}_{i}=\widehat{\phi}_{0i}Rd\eta_{i}/dt-\widehat{r}_{0i}%
R\omega_{0}\eta_{i}\label{nn3}%
\end{equation}
through first order in the perturbation $\eta_{i}.$ \ In obtaining Eqs.
(\ref{nn2}) and (\ref{nn3}), we have used Eqs. (\ref{n00a}) and (\ref{n00b})
as well as the small-angle approximations $\cos(\phi+\delta\phi)\approx
\cos\phi-\delta\phi\sin\phi$ and $\sin(\phi+\delta\phi)\approx\sin\phi
+\delta\phi\cos\phi.$ \ 

The mechanical momentum in Eq. (\ref{e2a}) involves two terms, $\mathbf{p}%
_{i}^{mech}=m_{i}\mathbf{v}_{i}+m_{i}v_{i}^{2}\mathbf{v}_{i}/(2c^{2})$. \ When
averaged in time, the first term vanishes, $\left\langle m_{i}\mathbf{v}%
_{i}\right\rangle =m_{i}\left\langle \mathbf{v}_{i}\right\rangle =0,$ since
for a stationary situation the time-average velocity vanishes. \ Consequently,
the average mechanical momentum in a stationary situation involves terms which
already contain a factor of $1/c^{2},$ $\left\langle \mathbf{p}_{i}%
^{mech}\right\rangle =\left\langle m_{i}v_{i}^{2}\mathbf{v}_{i}/(2c^{2}%
)\right\rangle ,~$ so that the velocity $\mathbf{v}_{i}$ needs to be
calculated only through nonrelativistic order.

\subsection{Calculation of the Perturbation by Energy Conservation}

For the one-moving-particle magnet (or $N$-non-interacting-moving-particle
magnet), it is convenient to obtain the perturbed phase $\eta_{i}(t)$ from
energy conservation. \ The centripetal forces of constraint do no work, and
hence the total energy (kinetic plus electrostatic particle energy) of the
particle (or of each particle of a non-interacting group) is conserved,
\begin{equation}
\frac{1}{2}mv_{0i}^{2}=\frac{1}{2}mv_{i}^{2}+(eq/x_{q}^{2})R\cos[\omega
_{0}t+\theta_{i}+\eta_{i}(t)]\label{n4}%
\end{equation}
where the last term is the potential energy $-e_{i}\mathbf{E}_{q}%
(0)\cdot\mathbf{r}_{i}$ of the magnet charge $e$ in the approximately uniform
electrostatic field of the external charge $q$. \ Now we are interested in the
behavior of the system through first order in the perturbation $eq/x_{q}^{2}.$
\ Thus we expand each of the terms on the right-hand side of Eq. (\ref{n4}).
\ Expanding the particle kinetic energy $mv^{2}/2,$ we have from Eqs.
(\ref{n00}), (\ref{n3}) and (\ref{nn3})%
\begin{equation}
\frac{1}{2}mv_{i}^{2}=\frac{1}{2}m(\mathbf{v}_{0i}+\delta\mathbf{v)}^{2}%
=\frac{1}{2}mv_{0i}^{2}+m\mathbf{v}_{0i}\cdot\delta\mathbf{v}_{i}=\frac{1}%
{2}mv_{0i}^{2}+m\omega_{0}R^{2}\frac{d\eta_{i}}{dt}.\label{n5a}%
\end{equation}
Also, the term involving the cosine in Eq. (\ref{n4}) is already first order
in the perturbation, and therefore we may drop the $\eta_{i}$ in the argument
of the cosine. \ The energy conservation equation then becomes%
\begin{equation}
\frac{1}{2}mv_{0i}^{2}=\frac{1}{2}mv_{0i}^{2}+m\omega_{0}R^{2}\frac{d\eta_{i}%
}{dt}+\frac{eq}{x_{q}^{2}}R\cos(\omega_{0}t+\theta_{i})\label{n6}%
\end{equation}
or%
\begin{equation}
\frac{d\eta_{i}}{dt}=-\frac{eq}{x_{q}^{2}m\omega_{0}R}\cos(\omega_{0}%
t+\theta_{i}).\label{n7}%
\end{equation}
Integrating once, we have the perturbing phase as%
\begin{equation}
\eta_{i}(t)=-\frac{eq}{x_{q}^{2}m\omega_{0}^{2}R}\sin(\omega_{0}t+\theta
_{i}).\label{n8}%
\end{equation}

\subsection{Mechanical Momentum of the Perturbed One-Moving-Particle Magnet}

Since the external charge $q$ is not moving, the canonical momentum
$\mathbf{p}_{e}$ given in Eq. (\ref{e2}) for the one-moving-particle magnet
consists entirely of mechanical momentum. \ We can now calculate the average
mechanical linear momentum of the charge $e$ which is moving in a circular
orbit. \ Through first order in the perturbing force $eq/x_{q}^{2}$ and second
order in $v_{0}/c,$ the mechanical momentum $\mathbf{p}_{m}^{mech}$ is given
from Eqs. (\ref{e2a}) and (\ref{n3}) by
\begin{align}
\mathbf{p}_{e} &  =\mathbf{p}_{m}^{mech}=m\gamma_{e}\mathbf{v}_{e}%
=m[1+v_{0}^{2}/(2c^{2})+\mathbf{v}_{0}\cdot\delta\mathbf{v}_{i}/c^{2}%
](\mathbf{v}_{0i}+\delta\mathbf{v}_{i})\nonumber\\
&  =m(\mathbf{v}_{0i}+\delta\mathbf{v}_{i})+[v_{0}^{2}/(2c^{2})](\mathbf{v}%
_{0i}+\delta\mathbf{v}_{i})+m(\mathbf{v}_{0}\cdot\delta\mathbf{v}%
_{i})\mathbf{v}_{0i}/c^{2}.\label{n9}%
\end{align}
Then averaging in time and noting that $\left\langle \mathbf{v}_{0i}%
\right\rangle =\left\langle \delta\mathbf{v}_{i}\right\rangle =0,$we have from
Eq. (\ref{n00}), (\ref{nn3}), (\ref{n8}) and (\ref{n9}),
\begin{align}
\mathbf{p}_{e}  & =\left\langle \frac{m(\mathbf{v}_{0}\cdot\delta
\mathbf{v}_{i})\mathbf{v}_{0i}}{c^{2}}\right\rangle =\left\langle
\frac{m\omega_{0}R^{2}}{c^{2}}\frac{d\eta_{i}}{dt}\mathbf{v}_{0i}\right\rangle
\nonumber\\
& =\left\langle \frac{m\omega_{0}R^{2}}{c^{2}}\left(  -\frac{eq}{x_{q}%
^{2}m\omega_{0}R}\cos(\omega_{0}t+\theta_{i})\right)  \omega_{0}%
R[-\widehat{x}\sin(\omega_{0}t+\theta_{i})+\widehat{y}\cos(\omega_{0}%
t+\theta_{i})]\right\rangle \nonumber\\
& =-\widehat{y}\frac{eq\omega_{0}R^{2}}{2c^{2}x_{q}^{2}}.\label{n10}%
\end{align}
This result is just the negative of Eq. (\ref{e15}) when $N=1.$ \ Thus the
average mechanical momentum (for the single charge $e$ of the
one-moving-particle magnet in the presence of the charge $q$) is equal in
magnitude and opposite in sign from the canonical momentum $\mathbf{p}_{q}$
corresponding to the familiar electromagnetic field momentum associated with
the electric field of the external charge $q$ and the magnetic field of the
magnet. \ Thus the total momentum of the system consisting of the charge $q$
and the one-moving-particle magnet indeed vanishes,
\begin{equation}
\left\langle \mathbf{P}\right\rangle =\left\langle \mathbf{p}_{q}\right\rangle
+\left\langle \mathbf{p}_{e}\right\rangle =\left\langle \mathbf{p}_{q}%
^{field}\right\rangle +\left\langle \mathbf{p}_{e}^{mech}\right\rangle
=\widehat{y}\frac{eq\omega_{0}R^{2}}{2x_{q}^{2}c^{2}}-\widehat{y}%
\frac{eq\omega_{0}R^{2}}{2x_{q}^{2}c^{2}}=0,\label{n11}%
\end{equation}
as required by the relativistic conservation law. \ In this case, there are no
nonelectromagnetic forces which introduce local power into the system of the
magnet and the point charge $q,$ and hence the total linear momentum of this
system must vanish as required by Eq. (\ref{PcoE}). \ On the other hand, if we
consider the magnet alone as our system, then the electric forces
$e\mathbf{E}_{q}(\mathbf{r}_{i})$ of the external charge $q$ are now external
forces on the magnet, and these external forces provide local power
$e\mathbf{E}_{q}(\mathbf{r}_{i})\cdot\mathbf{v}_{i}$ to charges of the magnet.
\ Thus as far as the magnet alone is concerned, the power delivery by the
external forces is
\begin{equation}
\frac{1}{c^{2}}\sum\limits_{i}\left(  e\mathbf{E}_{q}(\mathbf{r}_{i}%
)\cdot\mathbf{v}_{i}\right)  \mathbf{r}_{i}=\frac{1}{c^{2}}\sum\limits_{i}%
\left(  e\frac{-\widehat{x}}{x_{q}^{2}}\cdot\mathbf{v}_{i}\right)
\mathbf{r}_{i}=\frac{\widehat{y}Nqe\omega_{0}R^{2}}{2c^{2}x_{q}^{2}%
}=-\left\langle \mathbf{p}_{e}\right\rangle ,\label{n12}%
\end{equation}
which agrees exactly with the mechanical momentum in the magnet, as required
by the relativistic conservation law in the form (\ref{intm}). \ The linear
momentum $\left\langle \mathbf{p}_{e}\right\rangle $ given in Eq. (\ref{n10})
is what is identified in the literature as the \textquotedblleft hidden
mechanical momentum\textquotedblright\ in the magnet.

\section{Internal Electromagnetic Momentum in the
Two-Interacting--Moving-Particle Magnet}

\subsection{Improved Model of Two Moving Magnet Charges which Interact}

The physical magnets found in nature do not consist of a single charged
particle (or of non-interacting particles) sliding on a frictionless ring.
\ Hence we turn to a model consisting of particles which interact pair-wise
with a partner on the opposite side of the ring as an improvement over our
noninteracting-particle model for a magnet. \ Since each pair of particles
does not interact with any other pair, we may treat the magnet model as
involving two \textit{interacting} moving charged particles. \ The limit to
pairwise interactions is made here simply for ease of calculation; a
fully-interacting calculation is carried out elsewhere.\cite{ABforces} \ Thus
now we have two charges $e$ held by external centripetal forces of constraint
in a circular orbit of radius $R$ centered on the origin in the $xy$-plane
while a neutralizing charge $Q=-2e$ is located at the origin. \ The
calculation for the magnetic moment $\overrightarrow{\mu}$ and the field
momentum associated with the canonical momentum $\mathbf{p}_{q}$ follow as in
the calculations above with the results in Eqs. (\ref{e10}) and (\ref{e15})
corresponding to $N=2$. \ 

\subsection{Internal Electromagnetic Field Momentum}

The total linear momentum $\mathbf{P}\ $of our example involves not only the
electromagnetic field momentum $\mathbf{p}_{q}=\mathbf{p}_{q}^{field}$
associated with the canonical momentum of the stationary particle $q$ but also
the canonical momenta $\mathbf{p}_{e1}$ and $\mathbf{p}_{e2}$ associated with
the particles of the magnet. \ Since the charge $q$ is at rest, it has no
magnetic field, and hence it does not contribute to the canonical momentum of
$\mathbf{p}_{e1}$ or $\mathbf{p}_{e2}.$ \ However, the canonical momentum
$\mathbf{p}_{e1}$ of the first magnet particle includes both its mechanical
momentum and also the electromagnetic momentum associated with its own
electric field and the magnetic field of the other moving particle in the
magnet. \ Now the \textit{unperturbed} motion of the magnet charges given in
Eqs. (\ref{n0}) and (\ref{n00}) involves no average linear momentum because
the two charges are always moving with opposite velocities on opposite sides
of the circular orbit. \ However, the \textit{perturbed} motion will indeed
involve net linear momentum for the magnet particles. \ As soon as the
particles of the magnet have mutual interactions, the mechanical kinetic
energy changes (which provided the basis for the internal \textit{mechanical}
momentum in the one-moving-particle magnet) are suppressed as energy goes into
electrostatic energy of the interacting particles. \ The internal
\textit{mechanical} momentum of the magnet decreases because of the
electrostatic interactions, and internal \textit{electromagnetic} momentum
appears in the internal electromagnetic fields. \ We will illustrate this
situation explicitly for our example involving the
two-interacting--moving-particle magnet.

\subsection{Calculation of the Perturbation Using Nonrelativistic Forces\ }

In order to obtain the internal linear momentum $\mathbf{p}_{e1}+$
$\mathbf{p}_{e2}$ of the magnet in the presence of the external electric field
$\mathbf{E}_{q}$ due to the charge $q,$ we need to calculate the perturbed
motion of the particles $e_{1}$ and $e_{2}.$ \ The perturbed positions of the
two charges $e_{1}$ and $e_{2}$ of the magnet will be written as in Eq.
(\ref{n2}), where the unperturbed initial phases differ by $\pi,$ $\phi
_{01}-\phi_{02}=\theta_{1}-\theta_{2}=\pi,$ and where it is again assumed that
$\eta_{i}(t)$ is a small correction. \ The perturbed velocities of the charges
are as given in Eq. (\ref{n3}), and the accelerations then follow as
\begin{align}
\mathbf{a}_{i}  & =(\omega_{0}+d\eta_{i}/dt)^{2}R\{-\widehat{x}\cos[\omega
_{0}t+\theta_{i}+\eta_{i}(t)]-\widehat{y}\sin[\omega_{0}t+\theta_{i}+\eta
_{i}(t)]\}\nonumber\\
& +(d^{2}\eta_{i}/dt^{2})R\{-\widehat{x}\sin[\omega_{0}t+\theta_{i}+\eta
_{i}(t)]+\widehat{y}\cos[\omega_{0}t+\theta_{i}+\eta_{i}(t)]\}\nonumber\\
& =-\widehat{r}_{i}(\omega_{0}+d\eta_{i}/dt)^{2}R+\widehat{\phi}_{i}(d^{2}%
\eta_{i}/dt^{2})R\label{e16b}%
\end{align}
where $\widehat{r}_{i}$ and $\widehat{\phi}_{i}$ are given in Eqs. (\ref{n2a})
and (\ref{n3a}).

Since the magnet charges $e_{1}$ and $e_{2}$ are constrained to move in a
circular orbit, the perturbation of the charges is determined by the
tangential acceleration. \ For charge $e_{i},$ the equation of motion requires
only the electrostatic forces due to the stationary charge $q$ and the moving
charge $e_{j\neq i}.$ \ There are relativistic fields of order $1/c^{2}$ in
the tangential direction due to the moving charge $e_{i\neq j}$, but these
$1/c^{2}$-corrections will not contribute to average momentum of the magnet
because the expression (\ref{e2b}) for the electromagnetic momentum of the
magnet is itself of order $1/c^{2}.$ \ The radial forces are balanced by the
forces of constraint. \ Thus from Eq. (\ref{e16b}), we have for the
nonrelativistic equation of motion%
\begin{align}
m\mathbf{a}_{i}\cdot\widehat{\phi}_{i}  & \approx mR\frac{d^{2}\eta_{i}%
}{dt^{2}}\nonumber\\
& =\widehat{\phi}_{i}\cdot\lbrack e\mathbf{E}_{q}(\mathbf{r}_{i}%
,t)+e\mathbf{E}_{j\neq i}(\mathbf{r}_{i},t)]=\widehat{\phi}_{i}\cdot\left(
\frac{-\widehat{x}eq}{x_{q}^{2}}+\frac{e^{2}(\mathbf{r}_{i}-\mathbf{r}_{j\neq
i})}{\left\vert \mathbf{r}_{i}-\mathbf{r}_{j\neq i}\right\vert ^{3}}\right)
\nonumber\\
& =\left(  -\widehat{\phi}_{i}\cdot\widehat{x}\frac{eq}{x_{q}^{2}%
}-\widehat{\phi}_{i}\cdot\mathbf{r}_{j\neq i}\frac{e^{2}}{\left\vert
\mathbf{r}_{i}-\mathbf{r}_{j\neq i}\right\vert ^{3}}\right) \label{e16c}%
\end{align}
since $\widehat{\phi}_{i}\cdot\mathbf{r}_{i}=0,$ where the tangential unit
vector is given in Eq. (\ref{n3a}) while $\mathbf{r}_{j\neq i}=\widehat{x}%
R\cos[\omega_{0}t+\theta_{i}+\pi+\eta_{j\neq i}(t)]+\widehat{y}R\sin
[\omega_{0}t+\theta_{i}+\pi+\eta_{j\neq i}(t)].$ \ Then we have
\begin{align}
\widehat{\phi}_{i}\cdot\mathbf{r}_{j\neq i}  & =R\{-\sin(\omega_{0}%
t+\theta_{i}+\eta_{i})\cos[\omega_{0}t+\theta_{i}+\pi+\eta_{j\neq
i}(t)]\nonumber\\
& +\cos(\omega_{0}t+\theta_{i}+\eta_{i})\sin[\omega_{0}t+\theta_{i}+\pi
+\eta_{j\neq i}(t)]\}\nonumber\\
& =R\sin(\eta_{j\neq i}-\eta_{i}+\pi)=-R\sin(\eta_{j\neq i}-\eta_{i})\approx
R(\eta_{i}-\eta_{j\neq i})\label{e16d}%
\end{align}
where we have used the approximation $\sin\phi\approx\phi$ for small $\phi.$
\ The separation $\left\vert \mathbf{r}_{i}-\mathbf{r}_{j\neq i}\right\vert $
between the charges is second order in the perturbation $\eta,$ and so we may
write $\left\vert \mathbf{r}_{i}-\mathbf{r}_{j\neq i}\right\vert \approx2R$ in
Eq. (\ref{e16c}). \ Then the nonrelativistic equation of motion (\ref{e16c})
for the charge $e_{i}$ through first order in the perturbation produced by
$eq/x_{q}^{2}$ becomes
\begin{equation}
mR\frac{d^{2}\eta_{i}}{dt^{2}}=\left(  \frac{eq}{x_{q}^{2}}\sin(\omega
_{0}t+\theta_{i})-\frac{e^{2}}{(2R)^{3}}R(\eta_{i}-\eta_{j\neq i})\right)
.\label{e16f}%
\end{equation}
We notice that since $\phi_{0i}-\phi_{0j\neq i}=\theta_{i}-\theta_{j\neq
i}=\pi,$ the right-hand side of this equation (\ref{e16f}) is odd under the
interchange of the two particles. \ Thus for the steady-state situation, we
must have
\begin{equation}
\eta_{j\neq i}=-\eta_{i}.\label{e16ff}%
\end{equation}
Then the perturbation in the phase $\eta_{i}$ (in steady state) is given by
\begin{equation}
mR\frac{d^{2}\eta_{i}}{dt^{2}}=\left(  \frac{eq}{x_{q}^{2}}\sin(\omega
_{0}t+\theta_{i})-\frac{e^{2}}{(2R)^{2}}\eta_{i}\right) \label{e16g}%
\end{equation}
with a steady-state solution%
\begin{equation}
\eta_{i}(t)=\frac{eq}{x_{q}^{2}}\frac{\sin(\omega_{0}t+\theta_{i})}%
{[-m\omega_{0}^{2}R+e^{2}/(2R)^{2}]}.\label{e16h}%
\end{equation}

If we take the magnitude $e$ of the charges as small (so that we may neglect
the terms in $e^{2}$ involving interactions between the charges), $m\omega
_{0}^{2}R\gg e^{2}/(2R)^{2},$ then equation (\ref{e16h}) agrees exactly with
the one-particle-magnet result in Eq. (\ref{n8}). \ Thus we recover the
non-interacting particle result in the appropriate small-charge-$e$ limit.
\ On the other hand, if the magnitude $e$ of the charges becomes large,
$e^{2}/(2R)^{2}\gg m\omega_{0}^{2}R,$ then according to Eq. (\ref{e16h}) the
electrostatic interaction contribution can dominate the mechanical
contribution. \ Although both the small-$e$ and large-$e$ results are
accurate, the near-resonant situation $-m\omega_{0}^{2}R+e^{2}/(2R)^{2}%
\approx0$ is not allowed by the approximation used in the calculation, that
the perturbation $\eta_{i}(t)~$\ is small, $|\eta_{i}(t)|\ll1.$

\subsection{Internal Momentum in the Magnet}

The internal canonical momentum of the magnet in the presence of the external
charge $q$ is given by the sum over the canonical momenta $\mathbf{p}_{i}$ in
Eq. (\ref{e2}) where the sum includes only the two charges of the magnet in
our model, each with canonical momentum
\begin{equation}
\mathbf{p}_{ei}=\mathbf{p}_{ei}^{mech}+\mathbf{p}_{ei}^{field}=m\left(
1+\frac{\mathbf{v}_{i}^{2}}{2c^{2}}\right)  \mathbf{v}_{i}+\frac{e^{2}}%
{2c^{2}}\left(  \frac{\mathbf{v}_{j\neq i}}{|\mathbf{r}_{i}-\mathbf{r}_{j\neq
i}|}+\frac{\mathbf{v}_{j\neq i}\cdot(\mathbf{r}_{i}-\mathbf{r}_{j\neq
i})(\mathbf{r}_{i}-\mathbf{r}_{j\neq i})}{|\mathbf{r}_{i}-\mathbf{r}_{j\neq
i}|^{3}}\right)  .\label{e25}%
\end{equation}
When averaged in time, we expect equal momentum contributions from each
charge. \ The velocity $\mathbf{v}_{i}$ is given in Eq. (\ref{n3}) where
$\eta_{i}$ is given in Eq. (\ref{e16h}) and its time derivative is%
\begin{equation}
\frac{d\eta_{i}}{dt}=\omega_{0}\frac{eq}{x_{q}^{2}}\frac{\cos(\omega
_{0}t+\theta_{i})}{[-m\omega_{0}^{2}R+e^{2}/(2R)^{2}]}.\label{e26}%
\end{equation}

\subsubsection{Mechanical Linear Momentum of a Perturbed Magnet Charge}

Then the mechanical contribution $\mathbf{p}_{i}^{mech}$ to the linear
momentum is%
\begin{align}
\mathbf{p}_{ei}^{mech}  & =m\left(  1+\frac{v_{i}^{2}}{2c^{2}}\right)
\mathbf{v}_{i}=m\mathbf{v}_{i}+m\frac{(\mathbf{v}_{0i}+\delta\mathbf{v}%
_{i})^{2}}{2c^{2}}\mathbf{v}_{i}\approx m\mathbf{v}_{i}+m\frac{\mathbf{v}%
_{0i}^{2}}{2c^{2}}\mathbf{v}_{i}+m\frac{\mathbf{v}_{0i}\cdot\delta
\mathbf{v}_{i}}{c^{2}}\mathbf{v}_{0i}\nonumber\\
& =m\mathbf{v}_{i}+m\frac{\mathbf{v}_{0i}^{2}}{2c^{2}}\mathbf{v}_{i}+\frac
{m}{c^{2}}\omega_{0}\frac{d\eta_{i}}{dt}R^{2}\mathbf{v}_{0i}\label{g1}%
\end{align}
from Eqs. (\ref{n3}) and (\ref{nn3}). \ The average value of the velocity is
zero, $\left\langle \mathbf{v}_{i}\right\rangle =0,$ since the magnet-charge
interaction is assumed stationary. \ The required average for $\mathbf{p}%
_{i}^{mech}$ follows from Eqs. (\ref{n00}), (\ref{e26}), and (\ref{g1}) as
\begin{equation}
\left\langle \mathbf{p}_{ei}^{mech}\right\rangle =\left\langle \frac{m}{c^{2}%
}\omega_{0}\frac{d\eta_{i}}{dt}R^{2}\mathbf{v}_{0i}\right\rangle
=\widehat{y}\frac{m}{c^{2}}\omega_{0}^{2}R^{2}\frac{eq}{x_{q}^{2}}\frac
{\omega_{0}R/2}{[-m\omega_{0}^{2}R+e^{2}/(2R)^{2}]}.\label{g2}%
\end{equation}

\subsubsection{Electromagnetic Linear Momentum Associated with a Perturbed
Magnet Charge}

The electromagnetic contribution $\mathbf{p}_{i}^{field}$\ corresponds to
\begin{align}
\mathbf{p}_{ei}^{field}  & =\frac{e^{2}}{2c^{2}}\left(  \frac{\mathbf{v}%
_{j\neq i}}{|\mathbf{r}_{i}-\mathbf{r}_{j\neq i}|}+\frac{\mathbf{v}_{j\neq
i}\cdot(\mathbf{r}_{i}-\mathbf{r}_{j\neq i})(\mathbf{r}_{i}-\mathbf{r}_{j\neq
i})}{|\mathbf{r}_{i}-\mathbf{r}_{j\neq i}|^{3}}\right) \nonumber\\
& =\frac{e^{2}}{2c^{2}}\left(  \frac{\mathbf{v}_{j\neq i}}{|\mathbf{r}%
_{i}-\mathbf{r}_{j\neq i}|}+\frac{\mathbf{v}_{j\neq i}\cdot\mathbf{r}%
_{i}(\mathbf{r}_{i}-\mathbf{r}_{j\neq i})}{|\mathbf{r}_{i}-\mathbf{r}_{j\neq
i}|^{3}}\right) \label{e27b}%
\end{align}
since $\mathbf{v}_{j\neq i}\cdot\mathbf{r}_{j\neq i}=0.$ \ The denominator
will involve a distance $2R$ through first order in the perturbation. \ We
need first
\begin{equation}
\mathbf{v}_{j\neq i}\cdot\mathbf{r}_{i}=v_{j\neq i}\widehat{\phi}_{j\neq
i}\cdot\mathbf{r}_{i}\approx\omega_{0}R^{2}(-2\eta_{i})\label{e27c}%
\end{equation}
from Eqs. (\ref{e16d}) and (\ref{e16ff}). \ Then from Eqs. (\ref{n00a}),
(\ref{e16h}), and (\ref{e27c}), the time-average of $(\mathbf{v}_{j\neq
i}\cdot\mathbf{r}_{i})\mathbf{r}_{i}$ becomes
\begin{equation}
\left\langle (\mathbf{v}_{j\neq i}\cdot\mathbf{r}_{i})\mathbf{r}%
_{i}\right\rangle =\left\langle \omega_{0}R^{2}(-2\eta_{i})\mathbf{r}%
_{i0}\right\rangle =(-2\omega_{0}R^{2})\frac{eq}{x_{q}^{2}}\frac{1}%
{[-m\omega_{0}^{2}R+e^{2}/(2R)^{2}]}\widehat{y}\frac{R}{2},\label{e27d}%
\end{equation}
and similarly
\begin{equation}
\left\langle (\mathbf{v}_{j\neq i}\cdot\mathbf{r}_{i})\mathbf{r}_{j\neq
i}\right\rangle =(2\omega_{0}R^{2})\frac{eq}{x_{q}^{2}}\frac{1}{[-m\omega
_{0}^{2}R+e^{2}/(2R)^{2}]}\widehat{y}\frac{R}{2}.\label{e27e}%
\end{equation}
Then from Eq. (\ref{e27b}), the time-average electromagnetic contribution to
$\mathbf{p}_{ei}$ is
\begin{align}
\left\langle \mathbf{p}_{ei}^{field}\right\rangle  & =\frac{e^{2}}{2c^{2}%
}\left[  (-2\omega_{0}R^{2})\frac{eq}{x_{q}^{2}}\frac{2}{[-m\omega_{0}%
^{2}R+e^{2}/(2R)^{2}]}\widehat{y}\frac{R}{2}\left(  \frac{1}{(2R)^{3}}\right)
\right] \nonumber\\
& =-\widehat{y}\left(  \frac{e^{2}}{(2R)^{2}}\right)  \frac{\omega_{0}R^{2}%
}{2c^{2}}\frac{eq/x_{q}^{2}}{[-m\omega_{0}^{2}R+e^{2}/(2R)^{2}]}.\label{e27h}%
\end{align}
Adding the mechanical contribution in Eq. (\ref{g2}) and the electromagnetic
contribution in Eq. (\ref{e27h}), we find%
\begin{align}
\left\langle \mathbf{p}_{ei}\right\rangle  & =\left\langle \mathbf{p}%
_{ei}^{mech}\right\rangle +\left\langle \mathbf{p}_{ei}^{field}\right\rangle
=\widehat{y}\frac{m}{c^{2}}\omega_{0}^{2}R^{2}\frac{eq}{x_{q}^{2}}\frac
{\omega_{0}R/2}{[-m\omega_{0}^{2}R+e^{2}/(2R)^{2}]}\nonumber\\
& -\widehat{y}\left(  \frac{e^{2}}{(2R)^{2}}\right)  \frac{\omega_{0}R^{2}%
}{2c^{2}}\frac{eq/x_{q}^{2}}{[-m\omega_{0}^{2}R+e^{2}/(2R)^{2}]}\nonumber\\
& =-\frac{\widehat{y}qe\omega_{0}R^{2}}{2c^{2}x_{q}^{2}}.\label{e27ha}%
\end{align}
The two equal contributions $\left\langle \mathbf{p}_{ei}\right\rangle $ and
$\left\langle \mathbf{p}_{ej\neq i}\right\rangle $ from the two particles give
the canonical momentum $\left\langle \mathbf{p}_{magnet}\right\rangle $\ of
the magnet as
\begin{equation}
\left\langle \mathbf{p}_{magnet}\right\rangle =\left\langle \mathbf{p}%
_{ei\text{ }}\right\rangle +\left\langle \mathbf{p}_{ej\neq i\text{ }%
}\right\rangle =-2\frac{\widehat{y}qe\omega_{0}R^{2}}{2c^{2}x_{q}^{2}%
}.\label{e27g}%
\end{equation}
But then the canonical momentum $\left\langle \mathbf{p}_{magnet}\right\rangle
$\ of the two-particle magnet is equal in magnitude and opposite in sign
compared to the canonical momentum $\left\langle \mathbf{p}_{q}\right\rangle $
of the external charge $q$ corresponding to $N=2$ \ in Eq. (\ref{e15}), which
was equal to the familiar electromagnetic field momentum involving the
electric field due to $q$ and the magnetic field of the magnet. \ We see that
the relativistic conservation law (\ref{coE}) regarding the center of energy
is indeed satisfied, and that the total internal momentum of the system of the
magnet and the point charge indeed vanishes. \ 

\section{Discussion of Internal Electromagnetic Momentum}

The magnitudes of $\left\langle \mathbf{p}_{ei}^{mech}\right\rangle $ in Eq.
(\ref{g2}) and of $\left\langle \mathbf{p}_{ei}^{field}\right\rangle $ in Eq.
(\ref{e27h}) are in the ratio $m\omega_{0}^{2}R/\left[  e^{2}/(2R)^{2}\right]
.$ \ Thus in the limit of small value for the charge $e$ of the magnet
particles, the mechanical momentum in the magnet dominates the internal
electromagnetic momentum of the magnet. \ This mechanical momentum corresponds
to the \textquotedblleft hidden mechanical momentum\textquotedblright\ of the
textbooks and literature. \ However, in the opposite limit of large charge $e$
for the magnet particles, the internal electromagnetic momentum of the magnet
becomes large and the mechanical momentum becomes negligible. \ As more
particles of fixed mass $m$ and charge $e$ are added to the magnet while
keeping the magnetic moment $\overrightarrow{\mu}$ fixed, the average speed
$v_{0i}=\omega_{0}R$ of the current carriers becomes ever smaller so that the
mechanical internal momentum becomes insignificant compared to the internal
electromagnetic momentum of the magnet. \ Thus for any physical multiparticle
magnet with its enormous number of charge carriers, we expect that only the
internal electromagnetic momentum of the magnet needs to be considered. \ This
internal electromagnetic momentum in the magnet is equal in magnitude and
opposite in direction to the electromagnetic field momentum which is found in
the elementary textbook calculations involving a point charge and a steady
current. \ The negligible contribution of the particle mechanical momentum to
the magnet's internal momentum is analogous to the negligible contribution of
the particle kinetic energy to the magnet field energy related to the
self-inductance of a circuit\cite{Faraday} where the mass and the charge of
the charge carriers is never mentioned in the textbooks. \ \ The familiar
electromagnetic momentum involving the electric field of the external charge
and the magnetic field of the magnet appears in all the textbooks. \ However,
there does not seem to be any discussion of the internal electromagnetic
momentum between the electric and magnetic fields of the magnet particles
undergoing perturbed motion due to the electric field of the external point
charge. \ 

Coleman and Van Vleck\cite{CV} insisted that the \textquotedblleft hidden
momentum\textquotedblright\ in magnets of Shockley and James\cite{SJ} was
purely \textit{mechanical}. Other authors (such as Furry\cite{F}) have not
emphasized this aspect. \ The idea of internal momentum in magnets now appears
in the electromagnetism textbooks. \ Jackson has a problem\cite{J618} in his
graduate text which considers a point charge at the center of a toroidal
magnet. \ There are no external forces on the system, the center of energy is
not moving, and therefore the total linear momentum of the system must vanish.
\ The text correctly suggests that some internal momentum is present in the
magnet, but follows Coleman and Van Vleck in referring to this momentum as
\textquotedblleft mechanical.\textquotedblright\ \ Also, in his undergraduate
text, Griffiths has an example\cite{G} of a rectangular current loop in an
external electric field, and notes correctly that relativistic internal
momentum (hidden momentum) is present in the current loop. \ For
noninteracting charges, the internal momentum is indeed \textquotedblleft%
\ purely mechanical;\textquotedblright\ however, for interacting charges, the
situation would involve internal electromagnetic momentum in the current loop.

\section{Acknowledgement}

I wish to thank Professor David J. Griffiths for his many helpful comments and
suggestions which improved the clarity and accuracy of the present article. \


\begin{thebibliography}{99}
\bibitem {Gauss}We will use gaussian units throughout our discussion. \ When
treating relativistic aspects of electromagnetic theory, gaussian units are
far more natural than S.I.
\bibitem {J597}The calculation of $1/(4\pi c)%
{\textstyle\int}
d^{3}r\mathbf{E}_{i}\times\mathbf{B}_{j}$ for two point charges is essentially
the same as the calculation of the vector potential in the Coulomb gauge given
on page 597 of Jackson's text in Ref. 4.
\end{thebibliography}
\end{document}